\documentclass[twocolumn, trackchanges]{aastex7}

\usepackage{amsmath}
\usepackage{bm} 
\usepackage{units}
\usepackage{hyperref}
\usepackage{multirow}

\usepackage{CJKutf8} 

\shorttitle{Hierarchical Test of Lorentz Invariance }
\shortauthors{Du et al.}

\begin{document}
\begin{CJK}{UTF8}{gbsn}

\title{Hierarchical Test of Lorentz Invariance with Gamma-Ray Burst Spectral-Lag Measurements}

\author[0000-0002-0986-218X]{Shen-Shi Du(杜绅仕)}
\affiliation{Department of Physics, Faculty of Arts and Sciences, Beijing Normal University, Zhuhai 519087, China}
\email{duss.physics@gmail.com}

\author[0000-0002-1027-6655]{Yi Gong(龚易)}
\affiliation{School of Electrical and Electronic Engineering, Wuhan Polytechnic University,  Wuhan 430023, China}
\email{gongyi@whu.edu.cn}

\author[0000-0003-0162-2488]{Jun-Jie Wei(魏俊杰)}
\affiliation{Purple Mountain Observatory, Chinese Academy of Sciences, Nanjing 210023, China}
\affiliation{School of Astronomy and Space Sciences, University of Science and Technology of China, Hefei 230026, China}
\email{jjwei@pmo.ac.cn}

\author[0000-0002-5550-4017]{Zi-Ke Liu (刘子科)}
\affiliation{INPAC, Shanghai Key Laboratory for Particle Physics and Cosmology, School of Physics and Astronomy, Shanghai Jiao Tong University, Shanghai 200240, China}
\email{zkliu888@sjtu.edu.cn}

\author[0000-0002-3309-415X]{Zhi-Qiang You (尤志强)}
\affiliation{Institute for Gravitational Wave Astronomy, Henan Academy of Sciences,
No. 228 Chongshili, Zhengzhou, China}
\email{zhiqiang.you@hnas.ac.cn}

\author[0000-0002-1122-1146]{Yan-Zhi Meng}
\affiliation{ School of Science, Guangxi University of Science and Technology, Liuzhou, Guangxi 545006, China
}
\email{yzmeng2023@126.com}

\author[0000-0001-7049-6468]{Xing-Jiang Zhu (朱兴江)}
\affiliation{Department of Physics, Faculty of Arts and Sciences, Beijing Normal University, Zhuhai 519087, China}
\affiliation{Institute for Frontier in Astronomy and Astrophysics, Beijing Normal University, Beijing 102206, China}
\email[show]{zhuxj@bnu.edu.cn}

\begin{abstract}
Gamma-ray bursts (GRBs) are among the most potent probes of Lorentz invariance violation (LIV), offering direct constraints on the quantum gravity energy scale ($E_{\rm QG}$) based on observations of energy-dependent time lags. 
Individual GRBs with well-defined positive-to-negative lag transitions have been used to set lower limits on $E_{\rm QG}$, 
but they suffer from uncertainties of spectral-lag measurements and systematics due to theoretical modelling of each burst. 
Here, we combine observations of 32 GRBs with positive-to-negative lag transitions to derive a statistically robust constraint on $E_{\rm QG}$ through hierarchical Bayesian inference. 
We find that the dominant systematic uncertainty in LIV constraints arises from the intrinsic lag modeling. 
Accounting for this uncertainty with cubic spline interpolation, we derive robust limits of $E_{\rm QG,1} \ge 4.37 \times 10^{16}$~GeV for linear LIV and $E_{\rm QG,2} \ge 3.02 \times 10^{8}$~GeV for quadratic LIV. 
We find that the probability for LIV, i.e., $E_{\rm QG,1}$ being below the Planck scale, is estimated to be around 90\%, which we conclude as no significant evidence for LIV signatures in current GRB spectral lag observations.
Our hierarchical approach provides a rigorous statistical framework for future LIV searches and can be extended to incorporate multi-messenger observations. 

\end{abstract}


 \section{Introduction} \label{sec:sec1} 

Lorentz invariance is a cornerstone of fundamental physics, 
stating that the laws of physics are identical in all inertial frames, 
independent of the observer's orientation and velocity.
However, several quantum gravity (QG) theories attempting to unify 
general relativity and Standard Model of particle physics  
predict that Lorentz invariance may break at energies approaching the Planck scale, 
$E_{\mathrm{pl}} = \sqrt{\hbar c^5 / (2\pi G)} \approx 1.22 \times 10^{19}$ GeV \citep{2005LRR.....8....5M, 2013LRR....16....5A}. 
Astronomical observations of high-energy emissions from gamma-ray bursts (GRBs; \citealt{1998Natur.393..763A}), pulsars \citep{1999A&A...345L..32K}, 
and active galactic nuclei (AGNs; \citealt{1999PhRvL..83.2108B}) have served as sensitive probes of Lorentz invariance violation (LIV; see \citealt{2022hxga.book...82W} for a recent review). 
The advent of multi-messenger astronomy has further diversified the avenues for LIV searches, enabling complementary tests with high-energy neutrinos, cosmic rays, and gravitational waves (for reviews, see, e.g., \citealt{2024rpgt.book..433D, 2025CQGra..42c2001A}). 
Although predicted LIV signatures are highly suppressed in current accessible energy regime ($E\ll E_{\mathrm{pl}}$), 
they typically increase with photon energy and can accumulate to observable levels over cosmological distances.

GRBs possess three particular attributions: large cosmological distances, millisecond variability, and high-energy emissions. 
These properties establish GRBs as powerful probes for constraining QG energy scale ($E_{\mathrm{QG}}$) in the dispersive photon sector (e.g., \citealt{1998Natur.393..763A, 2005LRR.....8....5M, 
2009Sci...323.1688A, 
2009Natur.462..331A,
2013APh....43...50E, 2019PhRvD..99h3009E, 2017ApJ...834L..13W, 2022hxga.book...82W, 2024PhRvL.133g1501C}). 
Spectral lags$-$arrival time delays between photons of differing energies$-$are a commonly observed characteristic in GRB light curves \citep{2000ApJ...534..248N}. 
Although positive lags can be well explained by physical models such as the ``curvature effect'' \citep{2001ApJ...554L.163I} and the bulk acceleration of the emission zone \citep{2016ApJ...825...97U}, rarely observed negative lags still lack a definitive theoretical explanation.  
\cite{2019ApJ...882..115D} suggested that a high-energy cutoff spectrum exhibiting soft-to-hard spectral evolution can explain the negative lag turnover at $\sim20$ MeV observed in GRB~160625B (see Figure 7 therein). 
However, such a spectral shape coupled with soft-to-hard spectral evolution patterns at high-energy bands has only been observed in GRB 160625B.

In contrast, QG effects can naturally account for negative lags via interactions between propagating photons and spacetime ``foam'', inducing an energy-dependent speed of light, known as vacuum dispersion  \citep{1997IJMPA..12..607A}.  
Existing constraints of $E_{\mathrm{QG}}$ based on the vacuum dispersion time-of-flight measurements consist of following analyses: (i) arrival time intervals between individual high- and low-energy photons \citep{1998Natur.393..763A, 2009Sci...323.1688A, 2009Natur.462..331A}; (ii) fits to the observed spectral lags using a linear regression model with the slope being connected to LIV contributions and the intercept corresponding to source-intrinsic lags \citep{2006APh....25..402E}; and (iii) global fits to the spectral lag$-E$ relations extracted from high-resolution light curves across broad energy bands, which exhibit a positive-to-negative lag transition (PNT) in the high energy range \citep{2017ApJ...834L..13W, 2021ApJ...906....8D}. 
Method (iii) is the state-of-the-art, as it provides a reasonable formulation of intrinsic energy-dependent time lags in a burst and, simultaneously, searches for LIV across broad energy bands \citep{2017ApJ...834L..13W}. All the above approaches can only yield lower limits on $E_{\rm QG}$,
because the theoretical values of time lags must be lower than the observed values. 

However, LIV constraints derived from individual GRBs are limited at observable energies ($E \ll E_{\rm pl}$) due to the uncertainties of spectral-lag measurements and theoretical models. 
Crucially, the significant burst-to-burst heterogeneity in the sensitivity to LIV signatures, primarily arising from inherent disparities in redshift, time lag amplitude, and PNT photon energy \citep{2022ApJ...935...79L, 2023JHEAp..40...41P}, renders GRBs noisy probes of $E_{\rm QG}$. 
Therefore, we propose to combine an ensemble of GRBs to limit $E_{\mathrm{QG}}$ and derive statistically meaningful LIV constraints. 
To achieve this, we develop a hierarchical Bayesian framework \citep{2004AIPC..735..195L, 2010ApJ...725.2166H}. 
This method incorporates the likelihood functions of a sample of GRBs and forward-models the distribution of $E_{\mathrm{QG}}$ posteriors, while accounting for uncertainties of $E_{\rm QG}$ estimates in each burst through marginalization.

This paper is organized as follows. 
In Section~\ref{sec:sec2}, we provide details of the spectral-lag data. 
Section~\ref{sec:sec3.1} introduces the parameterized framework of LIV. 
Section~\ref{sec:sec3.2} describes the Bayesian analyses of a single burst. 
Section~\ref{sec:sec3.3} establishes a hierarchical Bayesian framework that takes the posterior samples of 32 GRBs as input to infer the $E_{\rm QG}$ distribution. 
Section~\ref{sec:sec4} presents the main results, and we summarize in Section~\ref{sec:sec5}. 
We adopt a flat $\Lambda$CDM cosmological model with $\Omega_{\rm M,0}=0.315$ ($\Omega_{\Lambda,0}= 1- \Omega_{\rm M,0}$) and $H_0=67.36$ km s$^{-1}$ Mpc$^{-1}$ \citep{2020A&A...641A...6P} throughout this work. 

\section{Data}\label{sec:sec2}
We use the spectral lag catalog from \citet{2022ApJ...935...79L} (hereafter L22), which comprises 32 long-duration GRBs exhibiting significant PNTs\footnote{The spectral lag data sets of those GRBs are publicly available at \url{https://github.com/Shen-Shi/GRB_LIV_Hierarchical.git}.}. These GRBs were detected by \textit{Fermi}’s
Gamma-ray Burst Monitor ($8~\text{keV} - 40~\text{MeV}$; \citealt{2020ApJ...893...46V}), with spectral lags calculated via cross-correlation function \citep{2012ApJ...748..132Z}. The settings
and methods of light curve extraction, lag calculation, and detailed information of these GRBs are given by L22 in their Table 1.

\section{Method}\label{sec:sec3}

\subsection{LIV-Induced Lag}\label{sec:sec3.1}
We assume QG effects are parameterized by Taylor-expanded modified vacuum dispersion \citep{1998Natur.393..763A}:
\begin{equation}\label{eq:eq1}
E^2 = p^2c^2 \left[1 - \sum_{n=1}^{\infty} s_{\pm}\left( \frac{E}{E_{\mathrm{QG},n}} \right)^n \right], E\ll E_{\mathrm{pl}}, 
\end{equation}
where $E_{\mathrm{QG},n}$ denotes the QG energy scale at order $n$, $s_{\pm}=\pm 1$ with $s_{\pm}=+1$ ($s_{\pm}=-1$) corresponds to subluminal (superluminal) scenario, and $p$ and $c$ are the photon's momentum and propagating speed in vacuum, respectively. 
This induces energy-dependent photon group velocities: 
\begin{equation}\label{eq:eq2}
    v(E) \equiv \frac{\partial E}{\partial p} = c\left[1- \sum_{n=1}^{\infty} s_{\pm}\frac{n+1}{2}\left( \frac{E}{E_{\mathrm{QG},n}} \right)^n \right].
\end{equation}
Adopting its $n$th-order expansion of the leading LIV term produces time lags \citep{2008JCAP...01..031J}:
\begin{equation}\label{eq:eq3}
    \tau_{\rm LIV}(E) \equiv t_{\mathrm{L}}-t_{\mathrm{H}} = -s_{\pm} \frac{n+1}{2}\frac{E^n_{\mathrm{H}}-E^n_{\mathrm{L}}}{E^n_{\mathrm{QG},n}}\mathcal{K}(z), 
\end{equation}
where
\begin{equation}\label{eq:eq4}
    \mathcal{K}(z) = \int_0^zdz' \frac{(1+z')^n}{H_0[\Omega_{\rm M,0}(1+z')^3 + \Omega_{\Lambda,0}]^{1/2}},  
\end{equation}
and $t_{\mathrm{L/H}}$ denotes the arrival time of photons with energy $E_{\mathrm{L/H}}$. 

For a subluminal scenario ($s_{\pm} = +1$), Equation~(\ref{eq:eq3}) produces negative spectral lags induced by QG effects, which is the case we consider throughout this work. 
We focus on the constraints on the leading LIV terms up to quadratic order ($n=2$) within this framework.

\subsection{Single-Burst Bayesian analysis}\label{sec:sec3.2} 
We first conduct a Bayesian analysis to obtain the $E_{\rm QG}$ posteriors from the spectral lags ($\tau_{\rm obs}$) observed across broad energy bands for each GRB. 
We directly fit the observed spectral lag$-E$ relations with significant PNTs, in which $\tau_{\rm obs}$ consists of LIV-induced and source-intrinsic ($\tau_{\rm int} $) time lags:
\begin{equation}
\tau_{\rm obs} = \tau_{\rm LIV} + \tau_{\rm int,z},
\end{equation}
where $\tau_{\rm int,z} = \tau_{\rm int}(1+z)$ corrects the effect of time dilation induced by the expanding Universe in the observer's frame. 
Our first approach to fit $\tau_{\rm int,z}$ is to use a smooth broken power-law (SBPL) function used by L22, i.e., 
\begin{equation}\label{eq:tau_int}
    \tau_{\mathrm{int,z}}(E)
= \xi 
\left( \frac{E - e}{E_b} \right)^{\alpha_1}
\left\{
    \frac{1}{2}
    \left[
        1 + 
        \left( \frac{E - e}{E_b} \right)^{\frac{1}{\mu}}
    \right]
\right\}^{(\alpha_2 - \alpha_1)\mu},
\end{equation}
where $e$ is the known median value of the lowest reference energy band for calculating the (relative) spectral lags of high-energy photons, $\xi$ is the normalization amplitude, $\alpha_1$ and $\alpha_2$ denote the slopes before and after the transition energy $E_{\rm b}$,  respectively, and the smoothness of this transition is controled by $\mu$. 
We also adopt a non-parametric approach using cubic spline that is implemented in $\tt{Python}$ package $\tt{Scipy}$ \citep{2020SciPy-NMeth}.  
The cubic-spline interpolation minimizes functional assumptions and helps mitigate systematic effects from intrinsic-lag modeling in individual bursts. 

According to Bayes' Theorem \citep{2008ConPh..49...71T}, the posterior probability of the model parameters of a single burst is given such that
\begin{equation}\label{eq:posterior}
    P_0(\bm{\theta}| \tau_{\rm obs}) = \frac{\mathcal{L}_0(\tau_{\rm obs}| \bm{\theta}) P_0(\bm{\theta}) }{\int d\bm{\theta} \mathcal{L}_0(\tau_{\rm obs}| \bm{\theta}) P_0(\bm{\theta}) },
\end{equation}
where the integration in the denominator represents the evidence (or marginal likelihood) and serves as a normalization constant. $\mathcal{L}_0(\tau_{\rm obs}| \bm{\theta})$ represents the likelihood function of spectral lags of a GRB. For the parametric approach, where $\bm \theta = \{\xi, \alpha_1, \alpha_2, \mu, E_{\rm b}, \log_{10}(E_{\rm QG}/\text{GeV})\}$, the log-likelihood is given by
\begin{equation}\label{eq:single-burst-likelihood}
\ln \mathcal{L}_0 =
\begin{cases}
-\infty, & \alpha_1 < \alpha_2, \\[6pt]
-\dfrac{1}{2}
\displaystyle\sum_{i=1}^{n}
\left[
\dfrac{\bigl(\tau_{\mathrm{obs},i} - \tau_{\mathrm{mod},i}(\boldsymbol{\theta})\bigr)^2}
      {\sigma_{\tau_{\mathrm{obs},i}}^{2} + \sigma_{\tau_{\mathrm{mod},i}}^{2}(\boldsymbol{\theta})}
\right.\\
\qquad\left.
+ \ln\!\Bigl(
2\pi\,
\bigl[
\sigma_{\tau_{\mathrm{obs},i}}^{2} + 
\sigma_{\tau_{\mathrm{mod},i}}^{2}(\boldsymbol{\theta})
\bigr]
\Bigr)
\right],
& \alpha_1 \ge \alpha_2.
\end{cases}
\end{equation}
where $n$ is the number of spectral-lag data points for a burst, 
$\tau_{\rm mod}$ represents the theoretical values of time lags contributed by intrinsic lags and possible LIV effects, 
$\sigma_{\tau_{\mathrm{obs}}}$ is the spectral-lag measurement uncertainty, and $\sigma_{\tau_{\mathrm{mod}}}$ is the theoretical modeling uncertainty that is propagated from Equations~(\ref{eq:eq3}) and (\ref{eq:tau_int}) via $\sigma_{\tau_{\mathrm{mod}}}=[\dot{\tau}_{\rm LIV}(E) + \dot{\tau}_{\rm int,z}(E)]\sigma_E$, with $\sigma_E$ being the statistical error of photon energy in each energy band. For cubic-spline fits, we use the Gaussian log-likelihood term of Equation~(\ref{eq:single-burst-likelihood}) as $\ln \mathcal{L}_0$, setting $\sigma_{\tau_{\mathrm{mod}}}=\dot{\tau}_{\rm LIV}(E)\sigma_E$ and $\theta = \log_{10}(E_{\rm QG}/\text{GeV})$. Throughout this work, we define $\log_{10}(E_{{\rm QG},n}/\text{GeV})$ as $\log_{10}E_{{\rm QG},n}$. $P_0({\bm \theta})$ is the prior distribution, which encodes the knowledge of adopted models before seeing any observational data. 
By fitting intrinsic lags with a SBPL function, L22 adopted uninformative priors for a single GRB (see Table 2 therein) and obtained the posterior samples of $\log_{10} E_{\rm QG}$ (hereafter Sample I)\footnote{In our previous work \citep{2021ApJ...906....8D}, we adopted a simple power-law function, which only parameterizes positive intrinsic lags. To leverage the more sophisticated SBPL function used by L22, which can model both positive and negative lags, we use their posterior samples of $\log_{10} E_{\rm QG}$ as the input for our hierarchical Bayesian analyses.}, which are shown with gray histograms in Figures~\ref{fig:FigC} and \ref{fig:FigD} in Appendix~\ref{sec:append1}. 
In the case of fitting intrinsic lags via cubic spline interpolation, we adopt the same prior distributions as those used by L22 for the Bayesian sampling of $\log_{10} E_{\rm QG}$; the posterior distributions of resulting posterior samples (hereafter Sample II) are plotted with green histograms in Figures~\ref{fig:FigC} and \ref{fig:FigD} for a comparison. 

\begin{table}[htbp]
\centering
\caption{Logarithmic Bayes factor ($\ln \text{BF}$) and associated $1\sigma$ uncertainty obtained from single-burst Bayesian analysis. 
The compared models are cubic spline+LIV versus SBPL+LIV. The values of $\ln \mathrm{BF}^{n=1}$ ($\ln \mathrm{BF}^{n=2}$) are derived under linear (quadratic) LIV scenario.}
\label{tab:lnbf_results}
\begin{tabular}{lcc}
\hline\hline
Name & $\ln \mathrm{BF}^{n=1}$ & $\ln \mathrm{BF}^{n=2}$ \\
\hline
GRB080916C & $3.69 \pm 0.13$ & $3.25 \pm 0.12$ \\
GRB081221 & $11.73 \pm 0.12$ & $11.81 \pm 0.12$ \\
GRB090328 & $15.72 \pm 0.11$ & $16.09 \pm 0.11$ \\
GRB090618 & $2.67 \pm 0.14$ & $3.00 \pm 0.14$ \\
GRB090926A & $16.00 \pm 0.16$ & $13.77 \pm 0.14$ \\
GRB091003A & $25.05 \pm 0.09$ & $24.90 \pm 0.08$ \\
GRB100728A & $-5.92 \pm 0.12$ & $-5.71 \pm 0.12$ \\
GRB120119A & $5.37 \pm 0.12$ & $7.21 \pm 0.12$ \\
GRB130427A & $29.04 \pm 0.18$ & $28.95 \pm 0.18$ \\
GRB130518A & $26.59 \pm 0.12$ & $29.11 \pm 0.13$ \\
GRB130925A & $-5.74 \pm 0.16$ & $-9.76 \pm 0.13$ \\
GRB131108A & $30.88 \pm 0.12$ & $31.08 \pm 0.12$ \\
GRB131231A & $38.93 \pm 0.17$ & $34.22 \pm 0.14$ \\
GRB140206A & $33.74 \pm 0.14$ & $31.77 \pm 0.13$ \\
GRB140508A & $12.73 \pm 0.10$ & $14.61 \pm 0.10$ \\
GRB141028A & $3.93 \pm 0.10$ & $3.92 \pm 0.10$ \\
GRB150314A & $48.10 \pm 0.13$ & $48.28 \pm 0.12$ \\
GRB150403A & $11.80 \pm 0.11$ & $13.34 \pm 0.11$ \\
GRB150514A & $20.76 \pm 0.12$ & $20.07 \pm 0.10$ \\
GRB150821A & $1.90 \pm 0.13$ & $2.59 \pm 0.13$ \\
GRB160509A & $16.88 \pm 0.13$ & $16.19 \pm 0.11$ \\
GRB160625B & $52.10 \pm 0.14$ & $59.43 \pm 0.17$ \\
GRB171010A & $4.08 \pm 0.17$ & $1.11 \pm 0.15$ \\
GRB180703A & $24.40 \pm 0.10$ & $24.06 \pm 0.11$ \\
GRB180720B & $19.33 \pm 0.14$ & $20.25 \pm 0.14$ \\
GRB190114C & $48.63 \pm 0.17$ & $54.28 \pm 0.18$ \\
GRB200613A & $4.74 \pm 0.11$ & $6.27 \pm 0.12$ \\
GRB200829A & $63.03 \pm 0.16$ & $64.56 \pm 0.16$ \\
GRB201216C & $11.65 \pm 0.13$ & $12.54 \pm 0.13$ \\
GRB210204A & $12.64 \pm 0.13$ & $12.91 \pm 0.12$ \\
GRB210610B & $-3.29 \pm 0.15$ & $-6.04 \pm 0.13$ \\
GRB210619B & $60.40 \pm 0.15$ & $61.15 \pm 0.14$ \\
\hline
\end{tabular}
\end{table}

For each burst, we compute the evidence in Equation~(\ref{eq:posterior}) with nested sampling techniques \citep{2020MNRAS.493.3132S}, and the Bayes factors are given by the ratio of evidences of two competing models. The Bayes factors listed in Table~\ref{tab:lnbf_results} show strong to decisive preference for cubic-spline interpolation over the SBPL for 28 GRBs; only three GRBs (GRB~100728A, GRB~130925A, and GRB~210610B) prefer the SBPL model. In a Bayesian sense, the cubic spline is better supported by the lag data and provides a more appropriate description of the intrinsic lags for most GRBs in our sample, thereby allowing for more robust constraints on LIV.

\begin{figure*}[htbp]
    \centering
    \begin{tabular}{c}
        \includegraphics[width=0.98\linewidth]{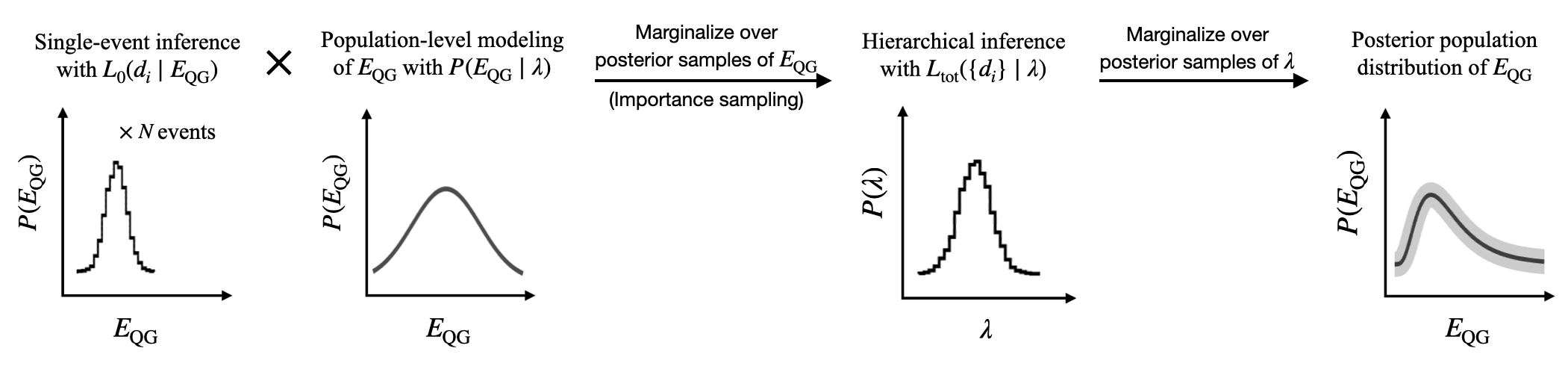} \\
    \end{tabular}
    \caption{
    Schematic illustration of the hierarchical Bayesian framework for inferring the distribution of $E_{\rm QG}$ from $N$ GRBs.
    For each burst, posterior samples of $E_{\rm QG}$ are inferred from the single-event likelihood $L_0(d_i \mid E_{\rm QG})$ as adopting a uniform prior with large range.
    The population-level distribution of $E_{\rm QG}$ is modeled by a probabilistic model $P(E_{\rm QG} \mid \lambda)$ parametrized by $\lambda$. Hierarchical inference is performed using the total likelihood $L_{\rm tot}(\{d_i\} \mid \lambda)$, which marginalizes the single-event likelihoods over the posterior samples of $E_{\rm QG}$. This marginalization can be evaluated by importance-sampling the posteriors of $E_{\rm QG}$. The posterior population distribution of $E_{\rm QG}$ is finally calculated by marginalizing over the posterior samples of $\lambda$.
    }
    \label{fig:Fig_method}
\end{figure*}

\begin{deluxetable*}{lcclclc}
\tablecaption{
The posterior samples used for hierarchical inference, hyper-parameters to be sampled, prior distributions, posterior credible intervals ($1\sigma$) for hyper-parameters, $e$-based logarithmic Bayes factors (Gaussian vs. log-normal), $1\sigma$ credible interval of $\log_{10} E_{\rm QG}$ obtained based on the mean values of PPD, and the probability of $E_{\rm QG,1} \leq E_{\rm pl}$ inferred from the mean values of PPD. 
\label{MyTabA}}
\small
\tablehead{
\colhead{Posterior sample} & 
\colhead{Parameter} & \colhead{Prior} & \colhead{Posterior($1\sigma$)} & ln BF & \colhead{$\log_{10} E_{\rm QG}$ ($1\sigma$)} & $P_{ E_{\rm QG,1}\leq  E_{\rm pl}}$ }
\startdata
Sample$-$I $\log_{10} E_{\rm QG,1}$ & $A$, $B$ & U(0,20),U(0,10) & $17.17^{+0.96}_{-0.93}$, $1.49^{+0.55}_{-0.55}$& 2.80 & [15.50, 18.54] &  92.14\%  \\
Sample$-$I $\log_{10} E_{\rm QG,1}$ & $C$, $D$ & U(0,20),U(0,10) & $17.00^{+0.87}_{-0.83}$, $0.09^{+0.03}_{-0.03}$ & $-$ & [15.46, 18.43] &  93.08\%  \\
Sample$-$II $\log_{10} E_{\rm QG,1}$ & $A$, $B$  &U(0,20),U(0,10)& $17.67^{+1.08}_{-0.90}$, $0.42^{+0.44}_{-0.30}$ & 2.70 & $\ge 16.64$ &  87.67\%  \\ 
Sample$-$II $\log_{10} E_{\rm QG,1}$ & $ C$, $ D$ &U(0,20),U(0,10)& $17.71^{+1.03}_{-1.00}$, $0.02^{+0.03}_{-0.02}$ & $-$ & $\ge 16.60$ &  87.82\%  \\
\hline
Sample$-$I $\log_{10} E_{\rm QG,2}$ & $A$, $B$  & U(0,15),U(0,10) & $7.35^{+0.74}_{-0.41}, 0.82^{+0.48}_{-0.30}$ & 1.72 & [6.53, 8.58] & $-$  \\
Sample$-$I $\log_{10} E_{\rm QG,2}$ & $C$, $D$  &U(0,15),U(0,10)& $7.36^{+0.62}_{-0.41}, 0.12^{+0.05}_{-0.04}$ & $-$ & [6.53, 8.62] & $-$  \\
Sample$-$II $\log_{10} E_{\rm QG,2}$ & $ A$, $B$ &U(0,15),U(0,10)& $10.47^{+1.68}_{-1.48}, 0.73^{+0.68}_{-0.50}$ & 2.48 & $\ge 8.48$ & $-$ \\
Sample$-$II $\log_{10} E_{\rm QG,2}$ & $C$, $ D$ &U(0,15),U(0,10)& $10.14^{+1.63}_{-1.44}, 0.07^{+0.07}_{-0.05}$ & $-$ & $\ge 8.72$ & $-$ \\
\enddata
\end{deluxetable*}

\begin{figure*}[htbp]
    \centering
    \begin{tabular}{cc}
        \includegraphics[width=0.42\linewidth]{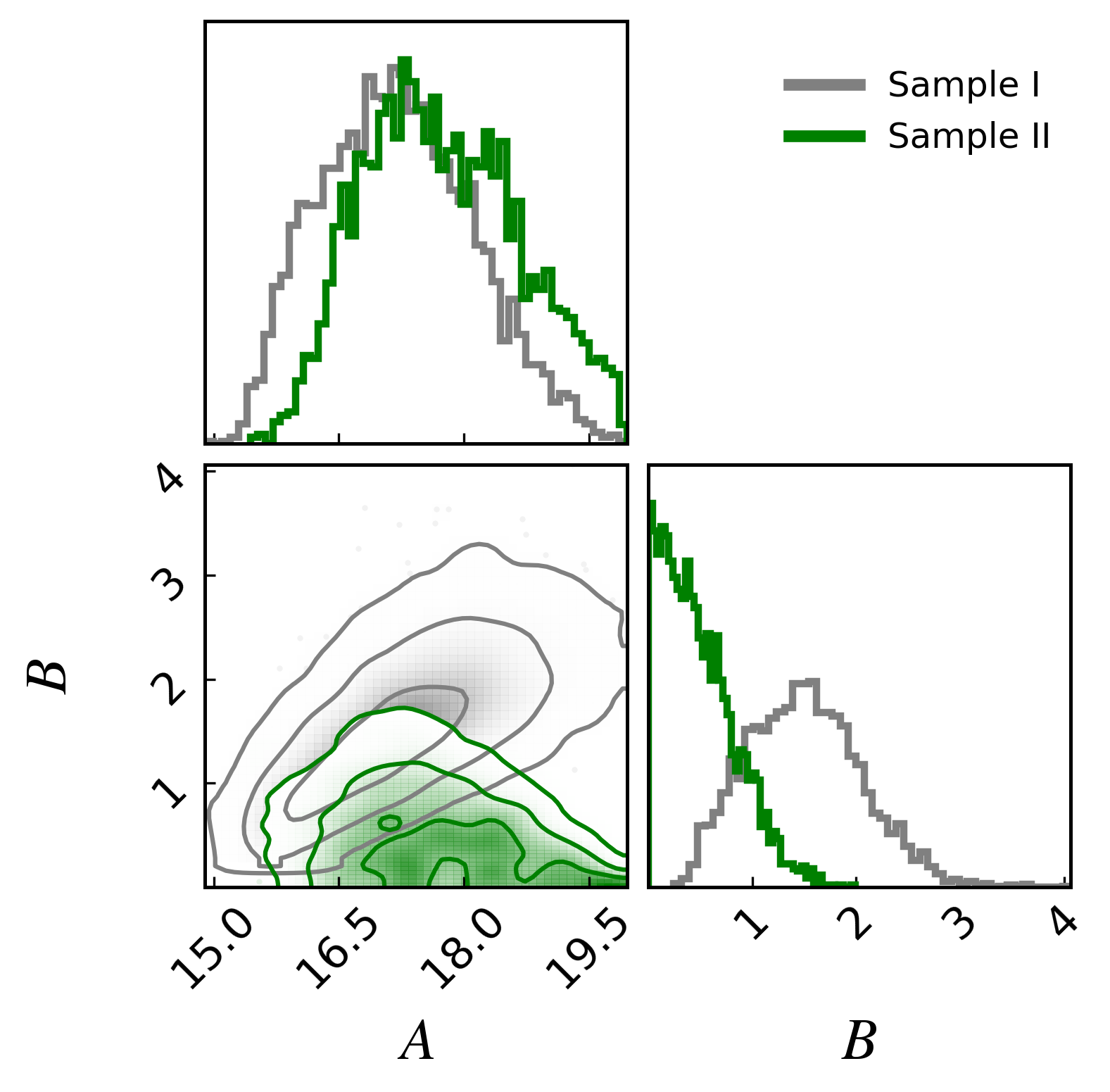} &
        \includegraphics[width=0.42\linewidth]{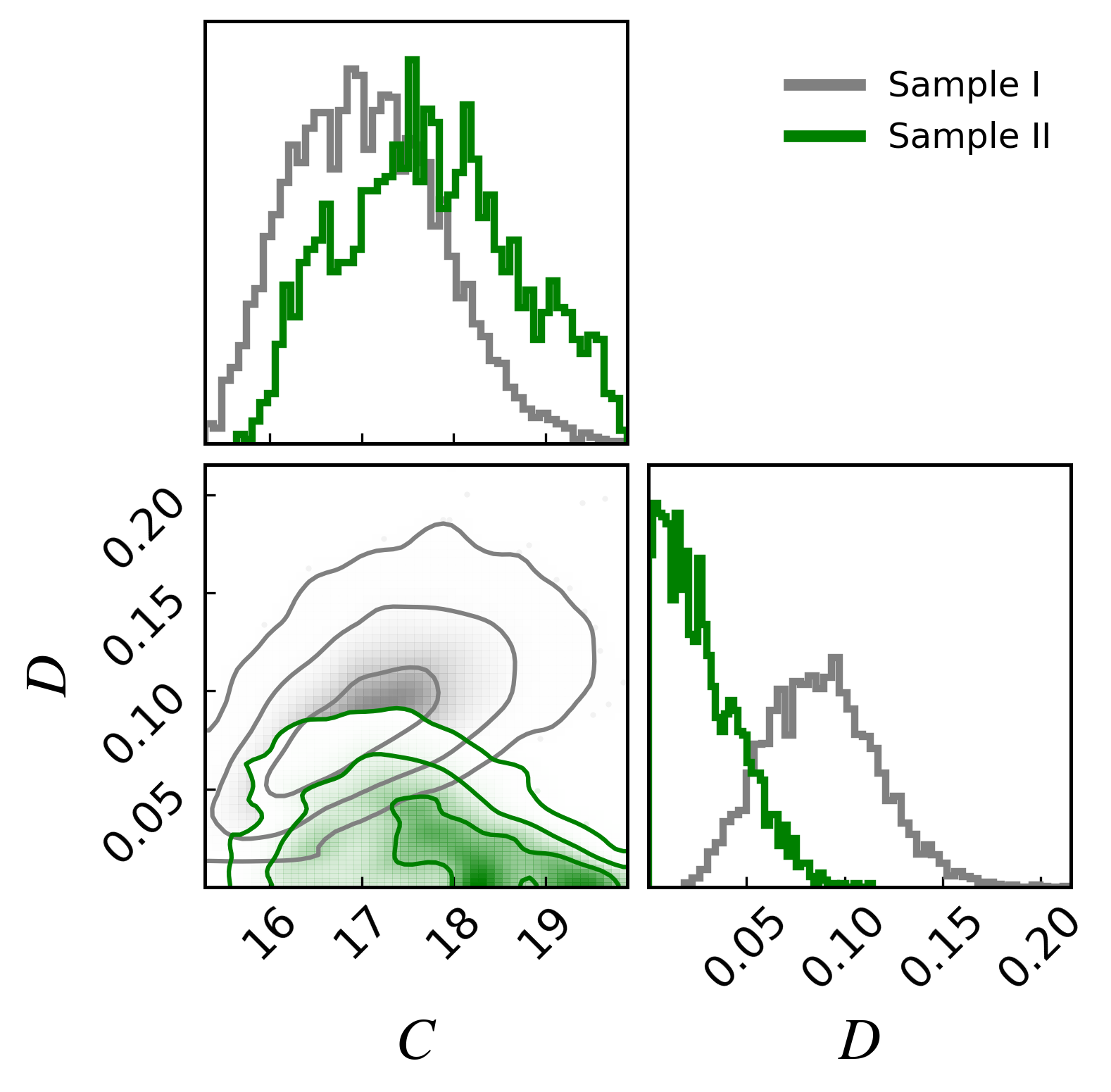} \\
        \includegraphics[width=0.42\linewidth]{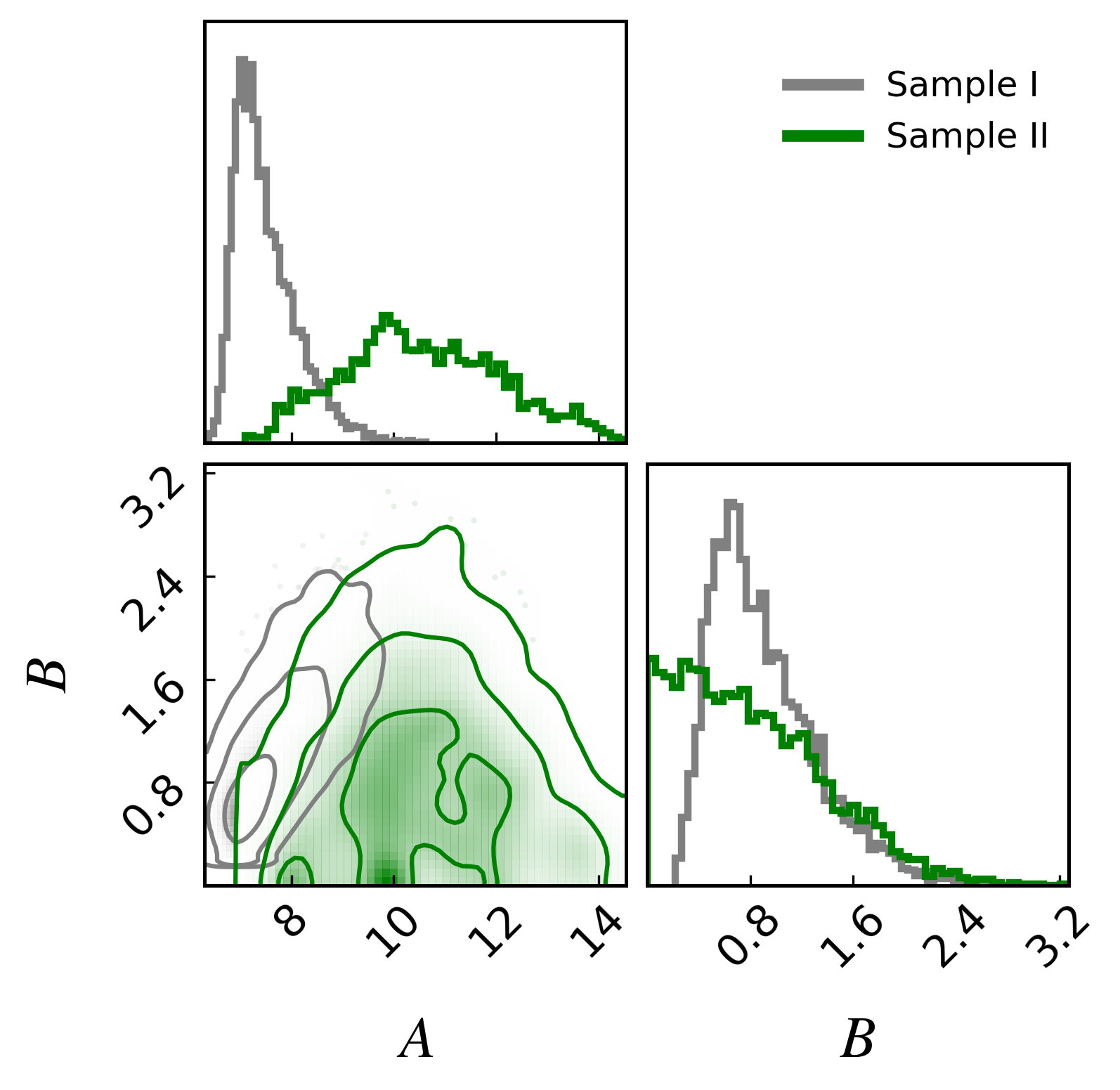} &
        \includegraphics[width=0.42\linewidth]{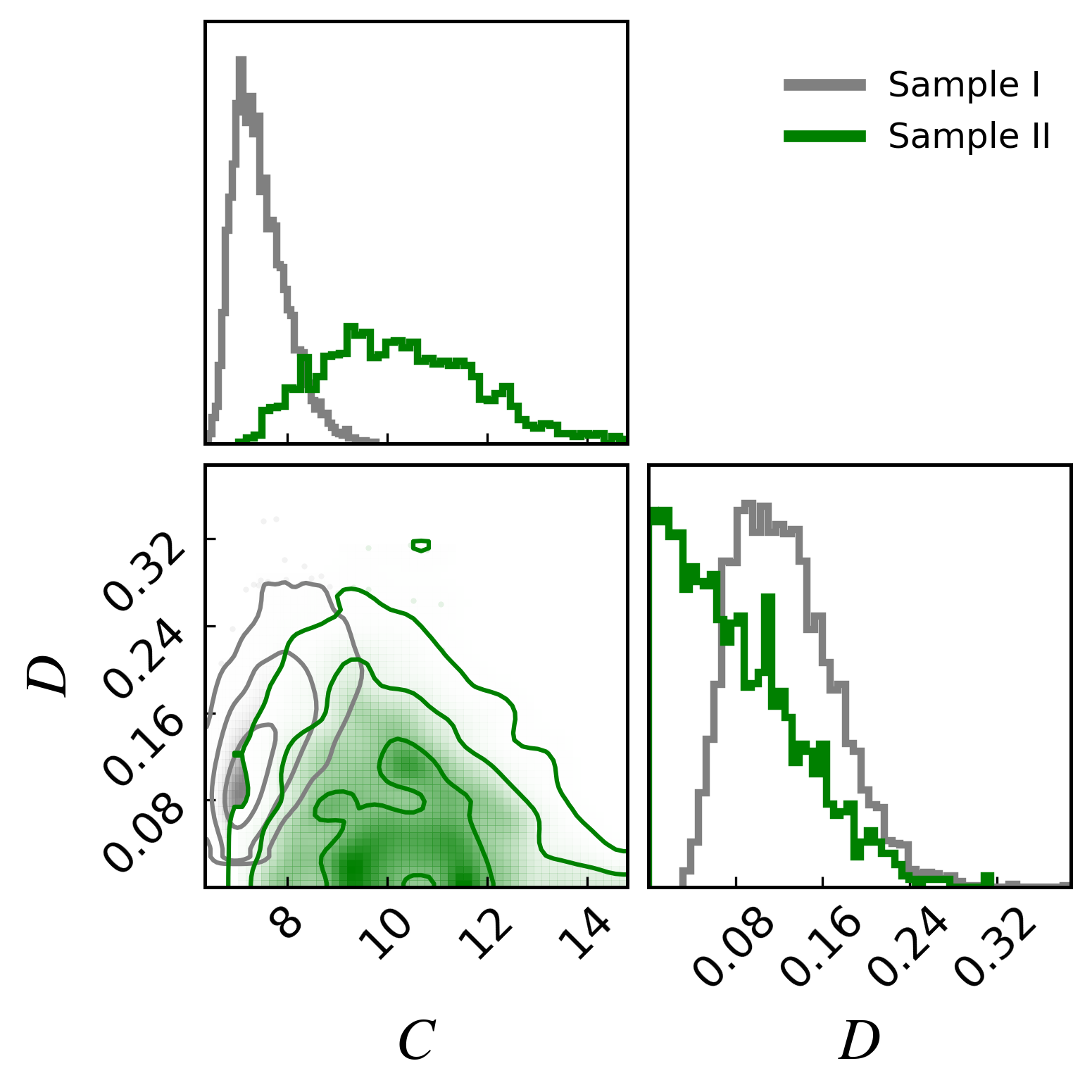} \\
    \end{tabular}
    \caption{One- and two-dimensional marginal posterior probability distributions of hyper-parameters. $A$ ($C$) and $B$ ($D$) are the parameters for Gaussian (log-normal) distribution model. 
    Upper and lower panels show the results  with analyzing the posterior samples of $\log_{10} E_{\rm QG,1}$ and $\log_{10} E_{\rm QG,2}$, respectively. In each panel, we compare the results inferred from Sample I and Sample II. }
    \label{fig:FigA}
\end{figure*}

\subsection{Hierarchical Bayesian  inference}\label{sec:sec3.3}
Given the posterior samples of $\log_{10} E_{\rm QG}$ inferred from $N=32$ GRBs, we seek to recover the $\log_{10} E_{\rm QG}$ distribution through forward modeling.  
Hierarchical Bayesian inference is an effective approach to combining an ensemble of samples to derive the distribution of parameters of interest, incorporating the different levels of measurement uncertainties of individual events \citep{2004AIPC..735..195L, 2010ApJ...725.2166H}. 
The total likelihood of all observed spectral lags $\{\tau_i\}$ is given by \citep{2010ApJ...725.2166H}:
\begin{equation}\label{eq:totallikelihood}
    \mathcal{L}_{\rm tot}(\{\tau_i\}| \lambda) = \prod_{i=1}^{N} \int d{\bm \theta}_i \mathcal{L}_0(\tau_i|{\bm \theta_i})P({\bm \theta_i}|\lambda),
\end{equation}
where
\begin{equation}\label{eq:population}
    P({\bm \theta}|\lambda) = \frac{ P(\log_{10} E_{\rm QG}| \lambda) P_0({\bm \theta})}{P_0(\log_{10} E_{\rm QG})},
\end{equation}
with $P(\log_{10} E_{\rm QG}| \lambda)$ being the $\log_{10} E_{\rm QG}$ distribution, which is parameterized by hyper-parameter(s), $\lambda$, we attempt to infer. Note that we have assumed that the spectral-lag observations of 32 GRBs are independent of each other, and the $\log_{10} E_{\rm QG}$ distribution is separable from other model parameters characterizing intrinsic time lags. 
The integration in Equation~(\ref{eq:totallikelihood}) accounts for the measurement uncertainties of each burst by marginalizing over the parameters for single-burst modeling.  
Inserting Equation~(\ref{eq:population}) into Equation~(\ref{eq:totallikelihood}), the total likelihood can be approximated by evaluating $\mathcal{L}_0$ through importance sampling over $K$ posterior samples derived from single-burst Bayesian analyses, that is
\begin{equation}\label{eq:importance-sampling}
    \mathcal{L}_{\rm tot}(\{\tau_i\}| \lambda) \approx \prod_{i=1}^{N} \frac{1}{K}\sum_{k=1}^{K} \frac{P(\log_{10} E_{{\rm QG,}ik}| \lambda)}{P_0 (\log_{10} E_{{\rm QG,}ik})}.
\end{equation}
Putting a prior for $\lambda$ results in the posterior probability as follows:
\begin{equation}\label{eq:eq10}
    P(\lambda| \{\tau_i\}) = \frac{\mathcal{L}_{\rm tot}(\{\tau_i\}| \lambda) P(\lambda)}{\int d\lambda \mathcal{L}_{\rm tot}(\{\tau_i\}| \lambda) P(\lambda)},
\end{equation}
where the denominator is the evidence of all involved GRB spectral lags.
The evidence can be used to perform model selection between two models $\lambda_1$ and $\lambda_2$ for the $\log_{10} E_{\rm QG}$ distribution by calculating the Bayes factor:
\begin{equation}
    \text{BF} =  \frac{\int d\lambda_2 \mathcal{L}_{\rm tot}(\{\tau\}| \lambda_2) P(\lambda_2)}{\int d\lambda_1 \mathcal{L}_{\rm tot}(\{\tau\}| \lambda_1) P(\lambda_1)}.
\end{equation}

\begin{figure*}[htbp]
    \centering
    \begin{tabular}{cc}
        \includegraphics[width=0.48\linewidth]{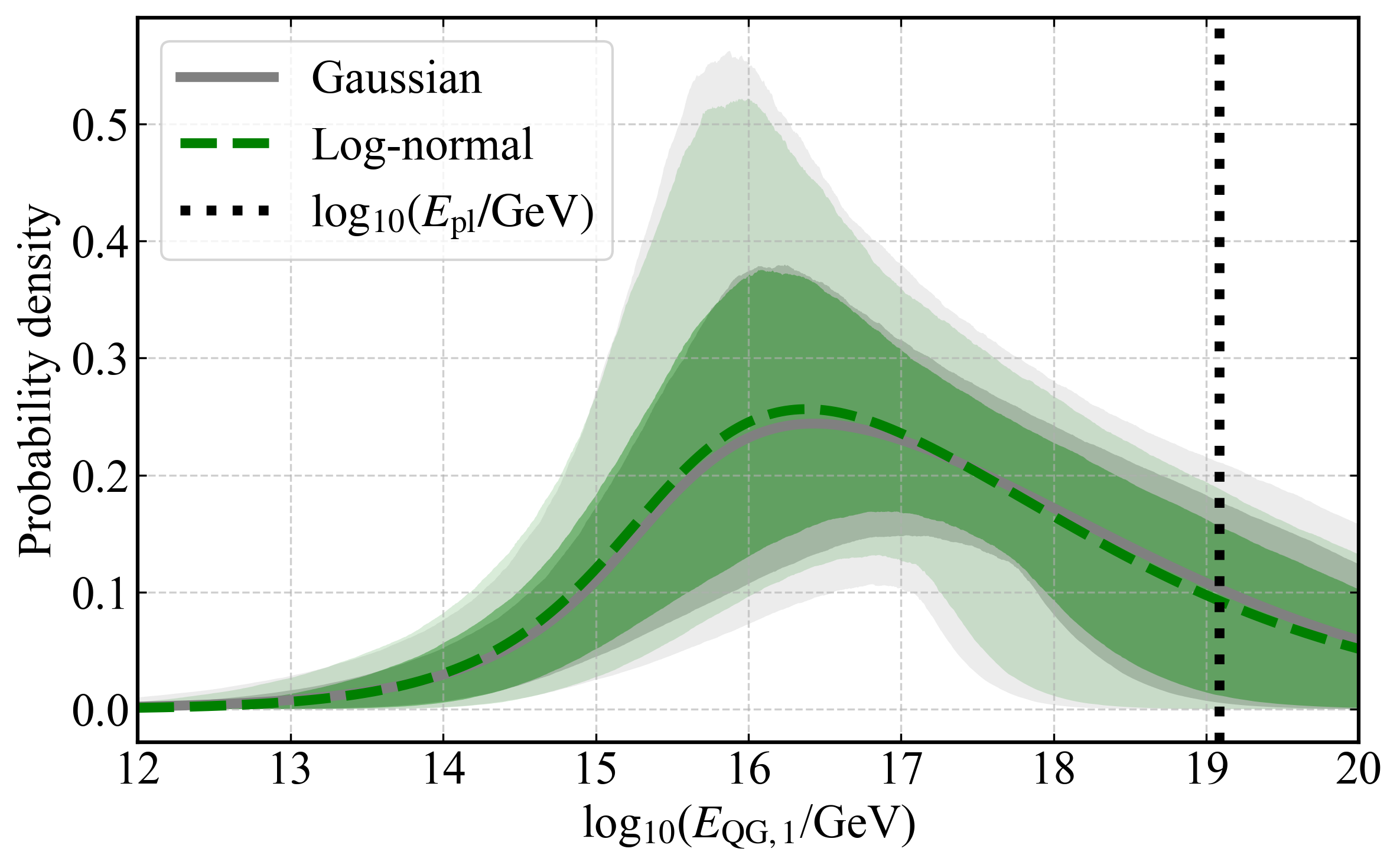} &
        \includegraphics[width=0.48\linewidth]{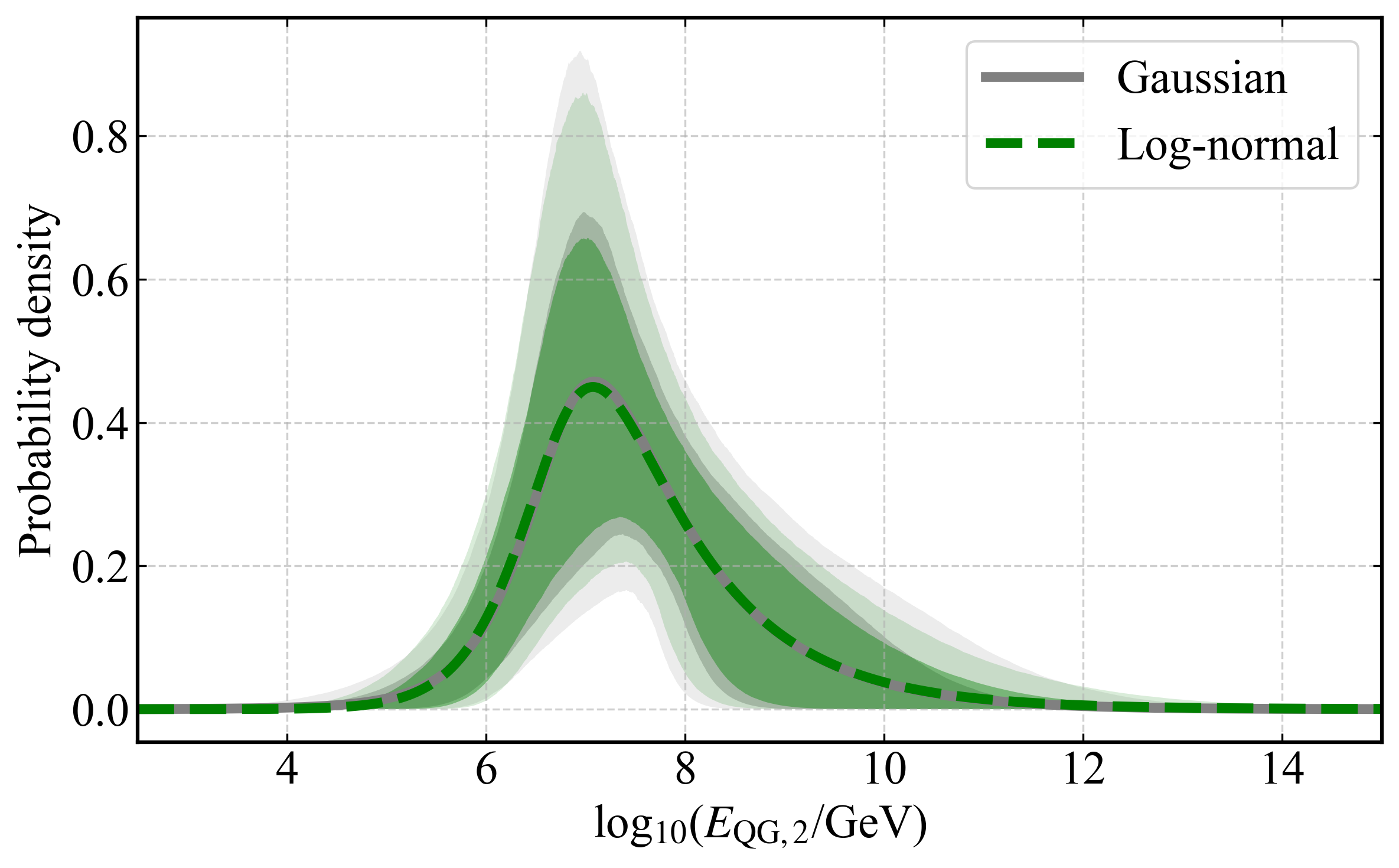} \\
        \includegraphics[width=0.48\linewidth]{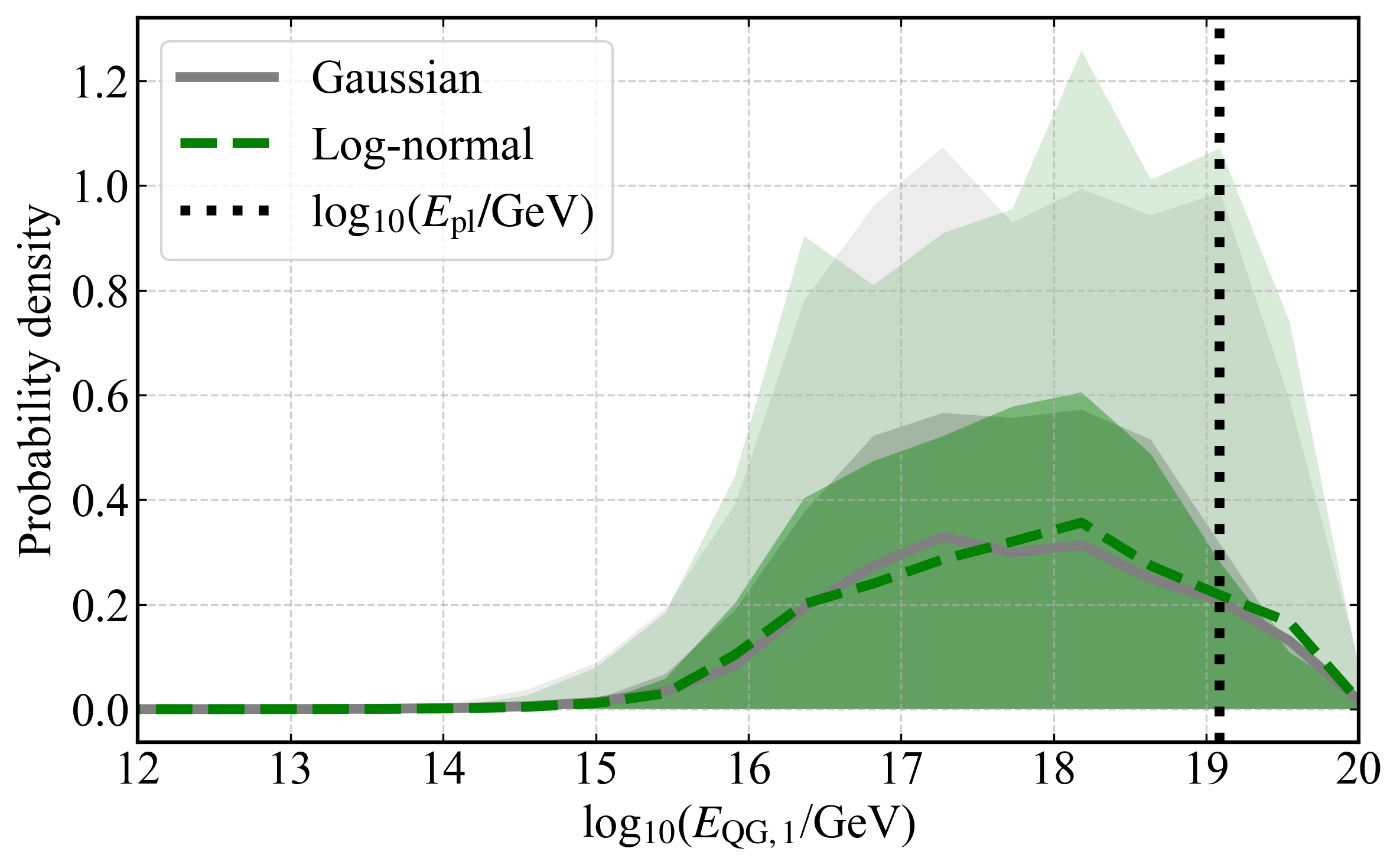} &
        \includegraphics[width=0.48\linewidth]{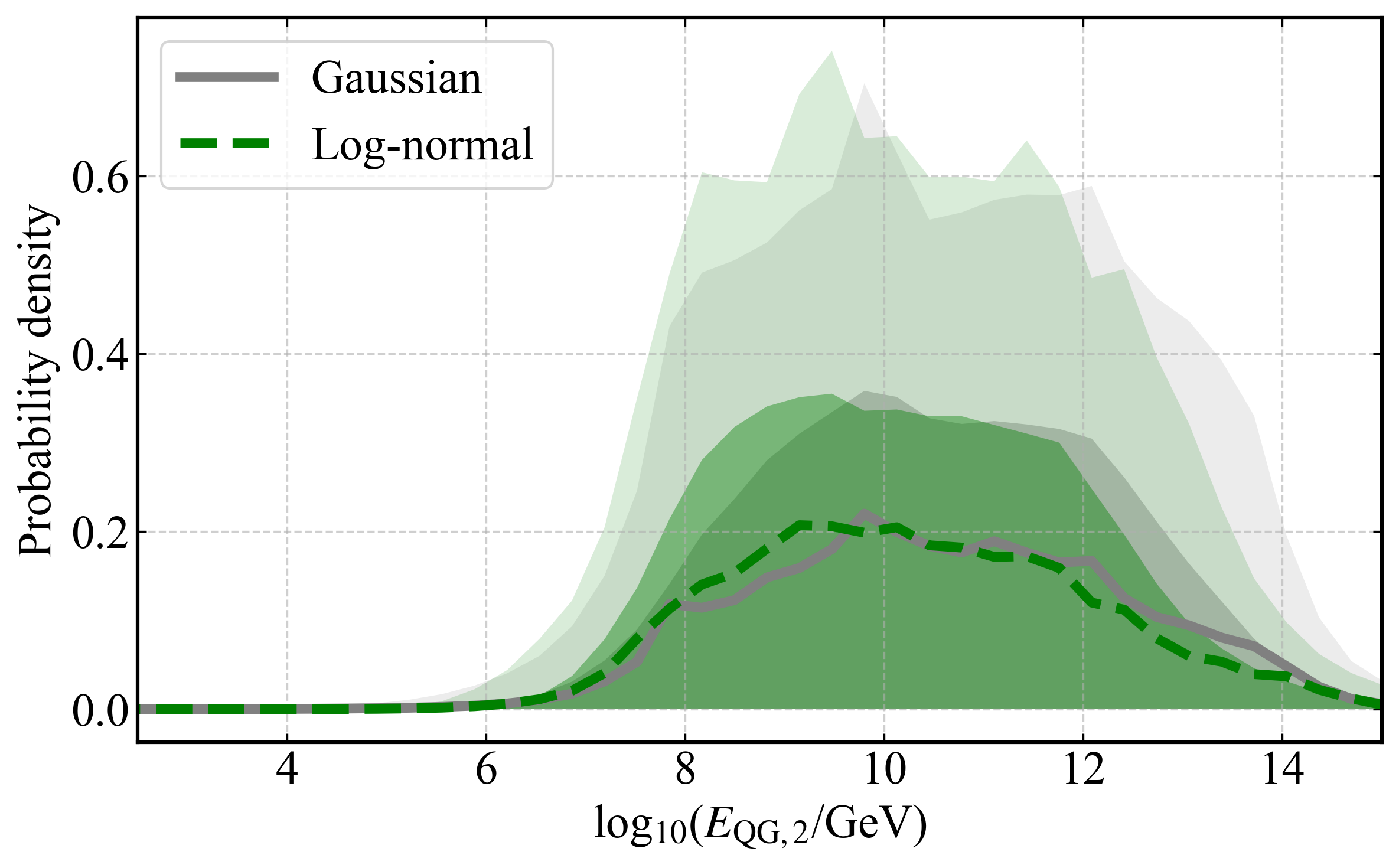} \\
    \end{tabular}
    \caption{Inferred posterior population distribution of quantum gravity energy scale. Top and bottom panels correspond to the results inferred from Sample I and Sample II, respectively. The mean PPDs are shown by solid and dashed lines, and the shaded areas present the $1\sigma$ (dark) and $2\sigma$ (light) credible intervals. The black dotted line denotes the Planck energy scale. }
    \label{fig:FigB}
\end{figure*}

Throughout this work, we assume two parametrized models for $P(\log_{10} E_{\rm QG}|\lambda)$, i.e., a Gaussian 
distribution
\begin{equation}\label{eq:gauss}
P(X |  A, B )
=\frac{1}{\sqrt{2 \pi} B} \exp \left[-\frac{1}{2}\left(\frac{X -A}{ B}\right)^{2}\right],
\end{equation}
and a log-normal distribution
\begin{equation}
P(X | C, D) = \frac{1}{\sqrt{2 \pi} D X } \exp \left[-\frac{\ln^2 (X/C)}{2D^2}\right],
\end{equation}
where $X=\log_{10} E_{\rm QG}$, $A$ and $B$ represent the mean and standard deviation of $\log_{10} E_{\rm QG}$ in the Gaussian model. In the log-normal distribution, $\ln(\log_{10} E_{\rm QG})$ follows a Gaussian distribution with mean $\ln C$ and standard deviation $D$; $ C$ and $ D$ provide the mean and standard deviation of $\log_{10} E_{\rm QG}$ via $C\cdot \exp(D^2/2)$ and $C\cdot \exp(D^2/2)\cdot \sqrt{\exp(D^2)-1}$, respectively. 
We opt to use these two probability distributions as they can explore the mean value and $1\sigma$ credible interval of $\log_{10} E_{\rm QG}$ that are inferred from the spectral lags of 32 GRBs. 
We use the $\tt{BILBY}$ \citep{bilby_paper} inference package with the nested sampling algorithm $\tt{DYNESTY}$ \citep{2020MNRAS.493.3132S} to numerically estimate the evidence in Equation~(\ref{eq:eq10}), while generating posterior samples for the parameters in each distribution model as by-products. We adopt uniform priors (denoted by U) with large ranges for hyper-parameters, as detailed in Table~\ref{MyTabA}.

After obtaining the posterior samples of $\lambda$, we perform posterior predictive checking by calculating the posterior population distribution (PPD):
\begin{equation}\label{eq:ppd}
    P(\log_{10}E_{\rm QG}| \{\tau_i\}) = \int d\lambda P(\lambda|\{\tau_i\}) P(\log_{10}E_{\rm QG}| \lambda). 
\end{equation}
This refers to the distribution of true $\log_{10}E_{\rm QG}$ values and is inferred from current spectral lag data by  marginalizing over $\lambda$ for a specific model.

In Figure~\ref{fig:Fig_method}, we illustrate the hierarchical Bayesian framework for inferring the distribution of $E_{\rm QG}$ using $N$ GRBs. The uncertainties in observed spectral lags and in the modeling of source-intrinsic and LIV-induced delays enter the single-burst likelihood [Equation~(\ref{eq:single-burst-likelihood})], and directly determine the width of the posterior distribution of $E_{\rm QG}$ for each GRB. Because the LIV-induced delay in the MeV energy bands is expected to be much smaller than the intrinsic lag, uncertainty in intrinsic-lag modeling dominates the uncertainty of $E_{\rm QG}$ constraints. These single-burst uncertainties are then propagated into the total likelihood [Equation~(\ref{eq:totallikelihood})] through marginalization over the posterior samples of $E_{\rm QG}$, thereby broadening the posterior distribution of the hyper-parameters $\lambda$. As illustrated in Figure~\ref{fig:Fig_method}, the final posterior population distribution of $E_{\rm QG}$ incorporates uncertainty in $\lambda$ through marginalization over its posterior samples, with the distribution width and shaded region explicitly presenting the overall uncertainties propagating from individual bursts to population inference of $E_{\rm QG}$.

\section{Results}\label{sec:sec4}
The one- and two-dimensional marginal posterior probability distributions of the hyper-parameters 
for both Gaussian and log-normal models are shown in Figure~\ref{fig:FigA}
(upper/lower panels for analyzing $\log_{10} E_{\rm QG,1}$/$\log_{10} E_{\rm QG,2}$), 
with summaries of results provided in Table~\ref{MyTabA}. 
In Figure~\ref{fig:FigB}, we show the PPD of $\log_{10}E_{\rm QG}$ 
derived from the posterior samples of hyper-parameters. Results are reported at the $1\sigma$ credibility unless otherwise stated.

We find the choice of the functional form for the $\log_{10} E_{\rm QG}$ distribution has a negligible effect on the recovered distribution, as both Gaussian and log-normal yield consistent mean values in all cases. 
Although statistically consistent, the resulting Bayes factors (Gaussian versus log-normal) indicate moderate evidence favoring the Gaussian, except for the quadratic LIV case with sample I, where we obtain weak evidence (see Table~\ref{MyTabA}).

In contrast, the intrinsic lag modeling in single-burst analyses 
significantly impacts the inference of $E_{\rm QG}$ distribution. 
Compared to the results of Sample I, using Sample II shifts the posterior probability distributions toward higher values of $E_{\rm QG}$ (Sample I versus Sample II in Figure~\ref{fig:FigA}), increasing the median values of $\log_{10} E_{\rm QG}$ by a factor of $\approx 5$ for linear LIV and $\approx 10^3$ for quadratic LIV (also see Table~\ref{MyTabA}). 
The shifts correlate with the enhanced flexibility for the intrinsic lag modeling to absorb energy-dependent time lags that would otherwise be attributed to LIV contributions.

The PPD of $\log_{10}E_{\rm QG}$ (Figure~\ref{fig:FigB}) provides the statistical results of the LIV constraints: 
\begin{itemize}
    \item With Sample I (top panels), the PPD peaks at $E_{\rm QG,1} \sim 10^{16}$ GeV for linear LIV and at $E_{\rm QG,2} \sim 10^{7}$ GeV for quadratic LIV. For the linear LIV, the PPD peak is around two orders of magnitude higher than the typical value reported by L22, while it is about one order of magnitude higher for the quadratic case. These typical values in L22 are obtained by fitting the histogram of the 1$\sigma$ lower bounds on $E_{\rm QG}$ with a Gaussian distribution (cf. their Figure~3). This discrepancy arises because the lower bounds of posterior distributions for each burst are lower than the peak or median values. As shown in Figure~\ref{fig:FigE} in Appendix~\ref{sec:append2}, when we analyze the lower bounds of Sample I using our method, the $E_{\rm QG}$ distribution peaks at $\sim 10^{14}$~GeV ($n=1$) and $\sim 10^6$~GeV ($n=2$), consistent with the results given by L22. 
    The credible intervals are ($3.16\times 10^{15}$, $3.47\times 10^{18}$)~GeV and ($2.88\times 10^{15}$, $2.69\times 10^{18}$) GeV in the linear LIV scenario under the Gaussian and log-normal models, respectively. For the quadratic case, the intervals are nearly identical for both distributions, spanning $E_{\rm QG,2}\approx 3\times 10^{6}$~GeV to $4\times 10^{8}$~GeV. 
    \item With Sample II (bottom panels), the constraints are more conservative. The Gaussian gives a lower limit of $E_{\rm QG,1} \ge 4.37 \times 10^{16}$ GeV for linear LIV and a weaker constraint of $E_{\rm QG,2} \ge 3.02 \times 10^{8}$ GeV for quadratic LIV. Log-normal results in comparable constraints: $E_{\rm QG,1} \ge 3.98 \times 10^{16}$ GeV and $E_{\rm QG,2} \ge 5.25 \times 10^{8}$ GeV.
\end{itemize}

Finally, we quantify the statistical evidence for LIV by computing the probability of $E_{\rm QG,1} \leq E_{\rm pl}$, based on the mean values of PPDs (solid and dashed lines in Figure~\ref{fig:FigB}).
Table~\ref{MyTabA} shows that this probability is 92.14\% for the Gaussian model and 93.04\% for the log-normal case by analyzing Sample I. It drops to 87.67\% and 87.82\%, respectively, for Sample II, well below the $3\sigma$ (99.73\%) credibility to claim evidence for LIV. 
We therefore do not find significant evidence for LIV signatures in our GRB spectral lag data.

\section{Summary}\label{sec:sec5}
In this work, we developed a hierarchical Bayesian framework to overcome the limitations of single-burst analyses in constraining quantum-gravity energy scale. 
By combining the posterior samples of $E_{\rm QG}$ derived from 32 GRBs with well-defined positive-to-negative spectral lag transitions, 
we derived statistically robust constraints on LIV. 
We performed hierarchical Bayesian analyses on two sets of posterior samples for $E_{\rm QG}$: Sample I, derived by fitting intrinsic time lags in individual GRBs with a smoothly broken power law, and Sample II, obtained with cubic spline interpolation that mitigates the systematic uncertainty associated with the functional form of intrinsic time lags.

Our main findings are summarized as follows:
\begin{itemize}
    \item With Sample I, the $E_{\rm QG}$ distribution peaks at $\sim 10^{16}$~GeV ($n=1$) and $\sim 10^7$~GeV ($n=2$), being roughly two orders of magnitude higher ($n = 1$) and about one order of magnitude higher ($n = 2$) than the typical values of $E_{\rm QG}$ reported by L22. The probability that $E_{\rm QG,1} \leq E_{\rm pl}$ is estimated to be 92.14\% (93.04\%) for the Gaussian (log-normal) distribution, well below the 99.73\% credibility threshold required to claim evidence for linear LIV. 
    \item By comparison, Sample II yields more conservative limits. Bayesian model selection moderately favors the Gaussian distribution over the log-normal in both the linear and quadratic LIV scenarios. Assuming the Gaussian model, we obtain a lower limit of $E_{\rm QG,1} \ge 4.37 \times 10^{16}$ GeV for linear LIV and $E_{\rm QG,2} \ge 3.02 \times 10^{8}$ GeV for quadratic LIV. The probability of linear LIV drops to 87.67\% (87.82\%) by assuming a Gaussian (log-normal) model. 
\end{itemize}
Our results suggest that the treatment of intrinsic lag modeling is the dominant systematic uncertainty in current LIV constraints based on the GRB observations of energy-dependent time delays.  

We conclude that the current GRB spectral lag data reveal no significant evidence for LIV signatures. 
Our approach establishes a statistically rigorous method for combining heterogeneous GRBs and multi-messenger probes of LIV. 
It is readily extendable to incorporate future GRB observations and to perform multi-messenger tests of Lorentz invariance with other astrophysical sources for a comprehensive search for new physics at the quantum-gravity energy scale.

\begin{acknowledgments}
We acknowledge the use of the OzSTAR supercomputing facility at Swinburne University of
Technology. 
Gong, Y. is supported by the Youth Talent Program of the Scientific Research Plan of the Hubei Provincial Department of Education (Grant No. Q20241713). 
Wei, J.-J. is supported by the Strategic Priority Research Program of the Chinese Academy
of Sciences (grant No. XDB0550400), the National Key R\&D Program of China (2024YFA1611704),
the National Natural Science Foundation of China (grant Nos. 12422307 and 12373053), and the Natural Science Foundation of Jiangsu Province (grant No. BK20221562). 
You, Z.-Q. was supported by the National Natural Science Foundation of China under Grant No. 12305059; The Startup Research Fund of Henan Academy of Sciences (Project No. 241841224); The Scientific and Technological Research Project of Henan Academy of Science (Project No. 20252345003); Joint Fund of Henan Province Science and Technology R\&D Program (Project No. 235200810111); Henan Province High-Level Talent Internationalization Cultivation Project(No.2024032).
Meng, Y.-Z. is supported by the Youth Program of National Natural Science Foundation of China (grant No. ‪12403045‬).
Zhu, X.-J. is supported by the National Natural Science Foundation of China (Grant No.~12203004), the National Key Research and Development Program of China (No. 2023YFC2206704), the Fundamental Research Funds for the Central Universities, and the Supplemental Funds for Major Scientific Research Projects of Beijing Normal University (Zhuhai) under Project ZHPT2025001.
\end{acknowledgments}

\vspace{5mm}

\software{ Bilby \citep{bilby_paper}, Dynesty \citep{2020MNRAS.493.3132S}, Scipy \citep{2020SciPy-NMeth}, Matplotlib \citep{Hunter:2007}, Numpy \citep{harris2020array}. }

\clearpage
\appendix

\section{Posterior distribution of $\log_{10} E_{\rm QG}$}
\label{sec:append1}

\begin{figure*}[htbp]
    \centering
    \begin{tabular}{c}
        \includegraphics[width=1\linewidth]{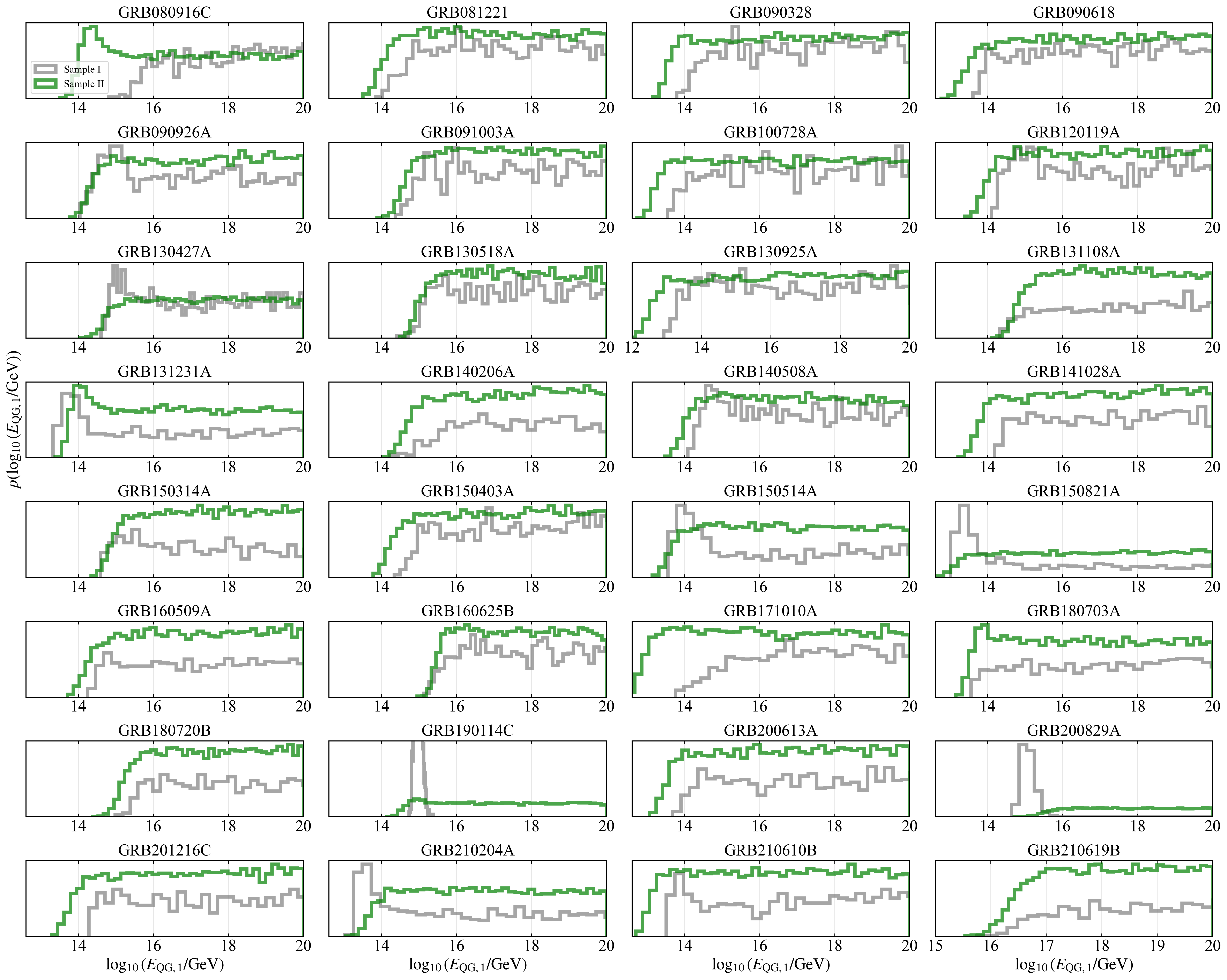} \\
    \end{tabular}
    \caption{Inferred marginal posterior distribution of linear quantum gravity energy scale from each GRB. Gray and green histograms correspond to  Sample I and Sample II, respectively. }
    \label{fig:FigC}
\end{figure*}

\begin{figure*}[htbp]
    \centering
    \begin{tabular}{c}
        \includegraphics[width=1\linewidth]{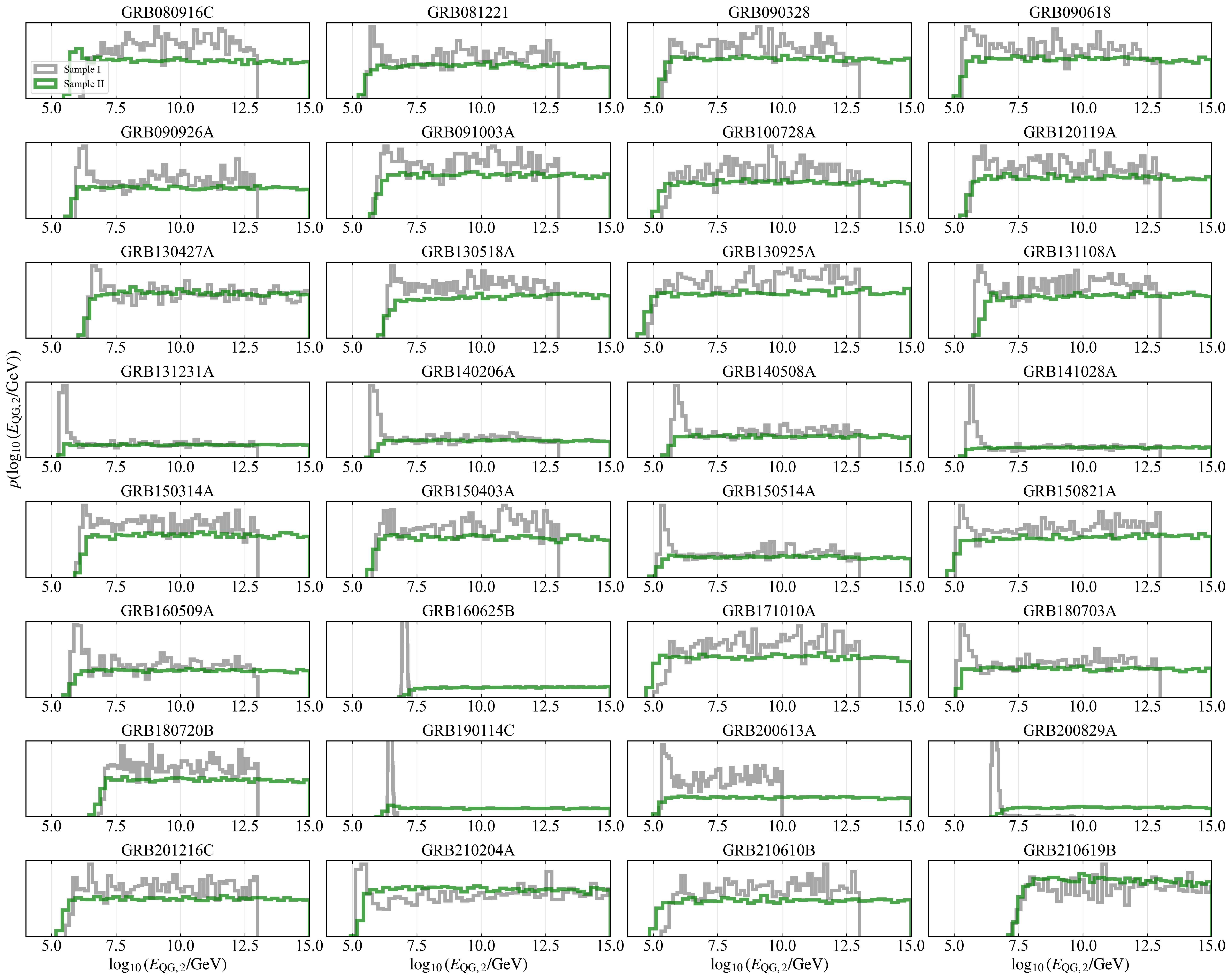} \\
    \end{tabular}
    \caption{Same as in Figure~\ref{fig:FigC}, but for the quadratic quantum gravity energy scale. }
    \label{fig:FigD}
\end{figure*}

Figures~\ref{fig:FigC} and \ref{fig:FigD} respectively display the posterior distributions of $\log_{10} E_{\rm QG,1}$ and $\log_{10} E_{\rm QG,2}$ derived from the Bayesian analyses of individual GRBs, as described in Section~\ref{sec:sec3.2}. 
In each panel, we plot the posterior distributions of the quantum gravity energy scale in Sample I (gray histogram) and Sample II (green histogram). 

\section{Comparison with L22's results}
\label{sec:append2}
\begin{figure*}[htbp]
    \centering
    \begin{tabular}{c}
        \includegraphics[width=0.5\linewidth]{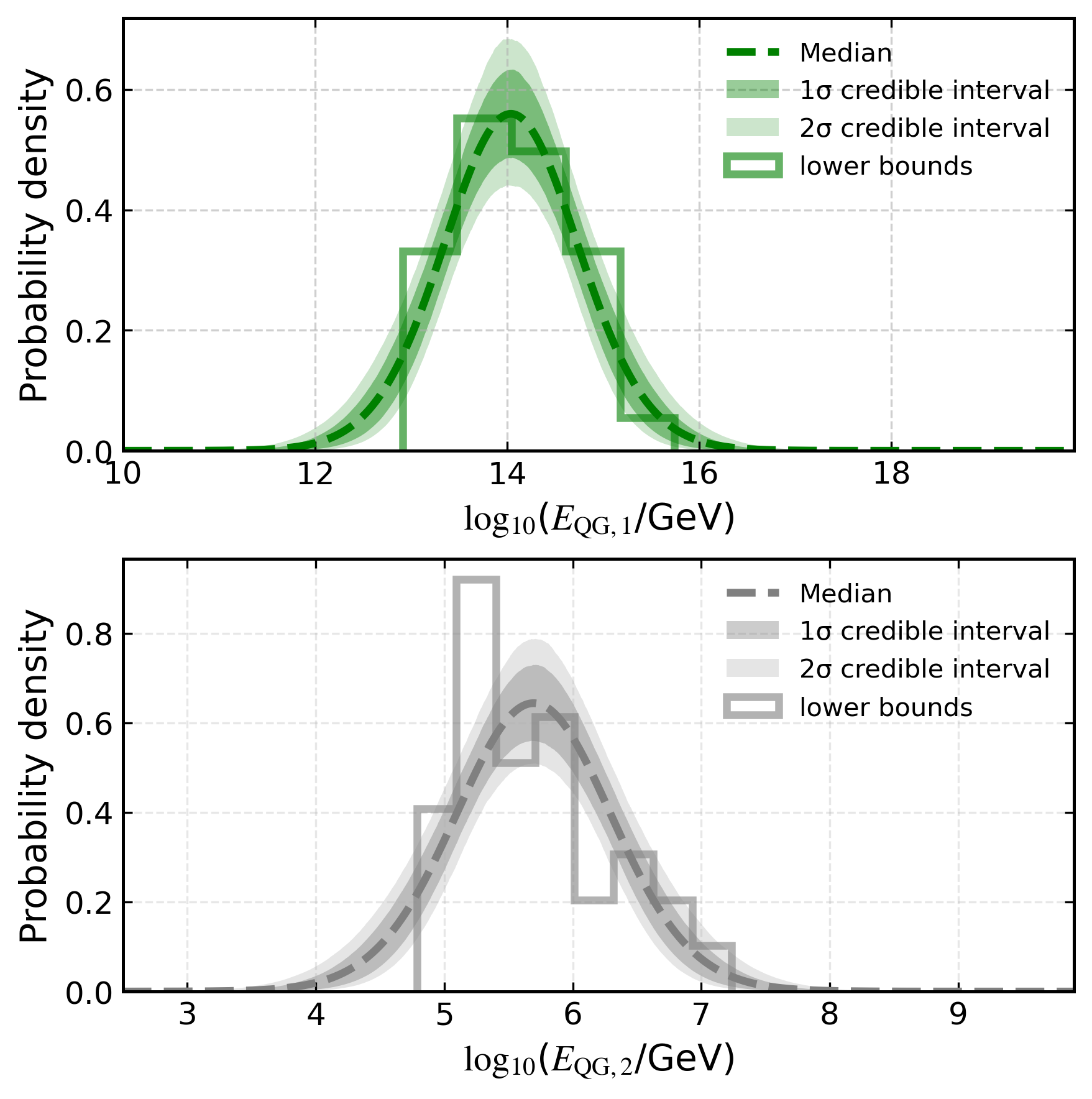} \\
    \end{tabular}
    \caption{The posterior population distribution of $\log_{10}E_{\rm QG}$ derived from the lower bounds of Sample I. }
    \label{fig:FigE}
\end{figure*}

To compare our results with those reported by L22, we perform hierarchical Bayesian inference on the $\log_{10}E_{\rm QG}$ distribution using only the lower bounds of Sample I as used by L22. 
For each burst, we draw 1000 posterior samples for $\log_{10}E_{\rm QG}$ from a \textit{Dirac} $\delta$ distribution centered at its lower limit; all samples are shown with the histograms in Figure~\ref{fig:FigE}. 
L22 fitted the histogram of lower limits (32 point estimates) with a Gaussian distribution and reported the mean values, so we use this model to fit the lower-bound posteriors for comparison. 
The PPDs are also shown in Figure~\ref{fig:FigE}, in which the upper and lower panels display the results for linear and quadratic LIV scenarios, respectively. 

\clearpage
\bibliography{cite}{}
\bibliographystyle{aasjournal}

\end{CJK}
\end{document}